\documentclass[letterpaper,10pt]{article} 
\usepackage{opticameet3} 

\usepackage{amsmath,amssymb}
\usepackage{cite}
\usepackage[colorlinks=true,bookmarks=false,citecolor=blue,urlcolor=blue]{hyperref} 
\usepackage{placeins}

\begin{document}

\title{Generalized Staircase Codes\\ with Arbitrary Bit Degree}

\author{Mohannad~Shehadeh, Frank~R.~Kschischang, and Alvin~Y.~Sukmadji}
\address{Department of Electrical and Computer Engineering,
University of Toronto, Toronto, Ontario, M5S 3G4, Canada}
\email{mshehadeh@ece.utoronto.ca, frank@ece.utoronto.ca, asukmadji@ece.utoronto.ca} 

\begin{abstract}
We introduce a natural generalization of staircase codes in which each bit is protected by arbitrarily many component codewords rather than two. This enables powerful energy-efficient FEC based on iterative decoding of Hamming components.
\end{abstract}

\section{Overview}

Staircase codes \cite{smith} and their generalizations 
\cite{sukmadji} have proven to be a highly-competitive paradigm
for high-rate, high-throughput, and high-reliability forward error
correction (FEC). 
Such codes use syndrome-domain iterative
bounded distance decoding of algebraic component
codes to enable high decoder throughputs
with relatively low internal decoder data-flow. 
Furthermore, their block-convolutional structures translate to natural 
pipelining and parallelism the combination of which is
essential to efficient hardware designs. These properties 
make them uniquely suited for next-generation FEC systems
which are characterized by terabit per second throughputs
and stringent energy-efficiency requirements.

Staircase codes and related product-like codes \cite{sukmadji}
typically protect each coded bit with two $t$-error-correcting
Bose--Chaudhuri--Hocquenghem (BCH) component codes. 
This typically requires that
$t \geq 3$ to achieve error floors below $10^{-15}$
\cite{Low-Rate-Hardware-Staircase} which imposes a limit
on our ability to improve the energy-efficiency of such 
schemes since the power consumption of a BCH decoder
circuit is roughly proportional to $t^2$ \cite{BCH-Power-Estimate}.
Motivated by this, we propose a natural, non-trivial generalization of staircase 
codes in which each bit is protected by an arbitrary number
of component codewords rather than only two. The number
of component codewords protecting each bit is termed
the \emph{bit degree}.
This allows us to weaken the component
codes down to $t = 1$ components while still achieving 
error floors below $10^{-15}$ by compensating with a higher bit degree. 
Reasoning heuristically, 
we expect such schemes to be more 
energy-efficient than 
those employing a smaller number of stronger components since
a given correctable error pattern can 
be corrected with less energy assuming a $t^2$ power estimate 
for the correction of an equal share $t$ of the errors by each component. 
We further anticipate improved
performance--complexity--latency tradeoffs in a wider variety 
of contexts in light of recent works \cite{sukmadji-irregular,Huawei-Higher-Degree} 
which accomplish such goals via new product-like codes
with bit degree increased from two to three.

\section{Code Construction}

As with classical staircase codes \cite{smith}, 
a code is an 
infinite sequence $B_0, B_1, B_2, \dots$ of $S\times S$
matrices satisfying certain constraints. 
Matrix entries are from the binary field $\mathbb{F}_2$ for concreteness 
but non-binary alphabets are straightforwardly accommodated. 
We assume zero-based indexing of matrices with the $(i,j)$th entry of a matrix $B$ denoted by $(B)_{(i,j)}$
and the index set $\{0,1,\dots,S-1\}$ denoted by $[S]$.
Next, we specify a \emph{memory parameter} $M$ along
with a collection of $M+1$ permutations $\pi_k\colon [S] \times [S] \longrightarrow [S] \times [S]$ indexed by $k \in [M+1]$.
In other words, for each $k \in [M+1]$, we have an invertible function $\pi_k$ which maps each 
$(i,j) \in [S] \times [S]$ to $\pi_k(i,j) \in [S] \times [S]$.
We further define $\Pi_k\colon \mathbb{F}_2^{S\times S} \longrightarrow \mathbb{F}_2^{S\times S}$ 
to be the invertible function mapping a matrix $B \in \mathbb{F}_2^{S\times S}$ to 
its permuted copy $\Pi_k(B) \in \mathbb{F}_2^{S\times S}$ according to $\pi_k$. In 
particular, the entries of $\Pi_k(B)$ are given by $(\Pi_k(B))_{(i,j)} = (B)_{\pi_k(i,j)}$.
Finally, we specify a set of $M + 1$ distinct 
integers $d_0 < d_1 < \dots < d_M$ termed a \emph{ruler}
and a linear, systematic $t$-error-correcting \emph{component code} 
$\mathcal{C} \subseteq \mathbb{F}_2^{(M+1)S}$ of length $(M+1)S$
and dimension $(M+1)S-r$ thus having $r$ parity bits. We can
without loss of generality assume that $d_0 = 0$ and that $\Pi_0(B) = B$ or, equivalently, 
that $\pi_0(i,j) = (i,j)$.

A generalized staircase code is 
then defined by the constraint that the rows of the  
$S\times (M+1)S$ matrix 
\begin{equation}\label{constraint-span}
	\begin{pmatrix}
		\Pi_M(B_{i-d_M}) & \Pi_{M-1}(B_{i-d_{M-1}}) & \cdots & \Pi_{1}(B_{i-d_1}) & B_i
	\end{pmatrix}
\end{equation}
belong to $\mathcal{C}$ for all integers $i \geq d_M$ 
and the initialization condition $B_0 = B_1 = \cdots = B_{d_M - 1} = 0_{S\times S}$
where $0_{S\times S}$ is the all-zero matrix. 
These first $d_M$ blocks need not be transmitted since they are
fixed and encoding can be performed recursively similarly to \cite{smith}.
In particular, an $S\times(S-r)$ block of information bits is
adjoined to $M$ permuted previously encoded blocks to produce 
an $S\times r$ block of parity bits which, when adjoined to the 
information block, produces a block $B_i$ such that
the rows of \eqref{constraint-span} are codewords of $\mathcal{C}$.
As such, each block $B_i$ contains $S-r$ information columns
and $r$ parity columns yielding a nominal 
or unterminated code rate of $R_\mathsf{nominal} = 1-r/S$.
While the construction so far
guarantees that each bit is
protected by $M+1$ component codewords,
we further require the property that distinct
component codewords intersect in at
most one bit as in classical staircase codes
\cite{smith}. This property
is essential to error floor performance
and in fact guarantees that the 
lowest weight of an uncorrectable
error pattern under iterative
decoding is at least $(M+1)t+1$. Achieving this property requires a careful
and non-trivial choice of the permutations and ruler defining 
the code in \eqref{constraint-span}. 

It can be shown that if the ruler is
a \emph{Golomb ruler} \cite[Part~VI:~Chapter~19]{Colbourn} of order $M+1$ and
the $M+1$ permutations of $[S]\times [S]$
define an \emph{$(M+1,S)$-net} \cite[Part~III:~Chapter~3]{Colbourn}, then
distinct component codewords will intersect in at
most one bit. Both Golomb rulers and nets are well-studied 
combinatorial objects found in the standard reference 
\cite{Colbourn} and both have been separately
applied to code design, e.g., in \cite{CSOC} and 
\cite{SJ-PG} respectively.
We begin by defining the first of these objects.
The ruler $0 = d_0 < d_1 < \dots < d_M$ is said
to be a \emph{Golomb ruler} of order $M+1$ and
length $d_M$ if all positive differences
$d_j-d_i$ with $j > i$ are distinct. 
While any Golomb ruler suffices for our purposes, an
\emph{optimal} Golomb ruler, which is one having
the smallest possible length $d_M$ for a given order $M+1$,
is preferable since this minimizes the encoding and decoding
memory requirements. Optimal Golomb rulers of order $M+1$ for $M \in \{1,2,3,4\}$
are given by $\{0,1\}$, $\{0,1,3\}$, $\{0,1,4,6\}$,
and $\{0,1,4,9,11\}$ with these and optimal rulers
of order up to $M+1 = 20$ given in \cite[Part~VI:~Chapter~19]{Colbourn}. 

Next, we must define $(M+1,S)$-nets. While a standard definition
is given in \cite[Part~III:~Chapter~3]{Colbourn}, it suffices for
our purposes to note that such an object is
equivalent to our $M+1$ permutations of
$[S]\times [S]$ with the added
property that for all distinct 
$k,k'\in [M+1]$, 
any row of $\Pi_{k}(B)$ 
has precisely a single element in common with any
row of $\Pi_{k'}(B)$.
A variety of algebraic constructions
of such permutations are possible and
we provide two such choices which are
valid for $M \leq \mathsf{lpf}(S)$ where $\mathsf{lpf}(S)$
denotes the least prime factor of $S$. 
We take $\pi_0(i,j) = (i,j)$ and 
define for $k \in \{1,2,\dots,M\}$
\begin{equation}\label{net-perms}
	\pi_k(i,j) = 
		(j, i+(k-1)j)
	\pmod S \text{.}
\end{equation}
An alternative option is to take 
\begin{equation}\label{net-perms-inv}
	\pi_k(i,j) = 
	\begin{pmatrix}
		-(k-1)i+j, (1-(k-1)^2)i + (k-1)j
	\end{pmatrix}
	\pmod S
\end{equation}
which has the advantage over \eqref{net-perms} of being
a family of involutions, i.e., having $\pi_k = \pi_k^{-1}$.
Note that for either choice \eqref{net-perms} or \eqref{net-perms-inv},
we recover classical staircase codes \cite{smith} in the case of $M = 1$ since
we get that $\pi_1(i,j) = (j,i)$ or equivalently $\Pi_1(B) = B^\mathsf{T}$.
However, we emphasize that for $M > 1$, we must 
choose $S$ to be such that $M\leq \mathsf{lpf}(S)$ 
otherwise an $(M+1,S)$-net is not formed
and the error floor performance suffers. If $S = p$ a prime number, then
this condition simply becomes that $M\leq S$. 
Other choices
for the permutations are possible with 
more flexibility in the admissible values of $S$ 
if needed
by considering the results in \cite[Part~III:~Chapter~3]{Colbourn}.

\section{Simulation Results}

We consider the use of the proposed
codes with single-error-correcting, double-error-detecting  
extended Hamming components
along with the permutations \eqref{net-perms-inv} 
and the aforementioned optimal Golomb rulers. 
The component code 
$\mathcal{C}$ of length $(M+1)S$ and dimension $(M+1)S-r$
is then obtained by shortening a parent extended Hamming
code of length $2^{r-1}$ where $r-1 = \lceil \log_2((M+1)S) \rceil$.
We perform sliding window decoding of 
$W$ consecutive blocks at a time  where one iteration 
comprises decoding all constraints in the window consecutively
similarly to \cite{smith}. 
To facilitate simulation of the waterfall regime 
where error propagation events dominate, 
we convert to a
block code via a termination-like strategy in which a
frame is formed from $F$ consecutive blocks. For the
last $W$ of these $F$ blocks, the information
portion is presumed to be zero and only the parity
portion is transmitted with the encoder and decoder state
being reset between consecutive frames. This results in a 
block code rate of 
$R = ((S-r)(F-W))/(S(F-W)+Wr)$ 
which approaches $R_\mathsf{nominal} = 1-r/S$ as $F$ tends to infinity.

\begin{figure}[htbp]
  \centering
  \includegraphics[width=0.847\textwidth]{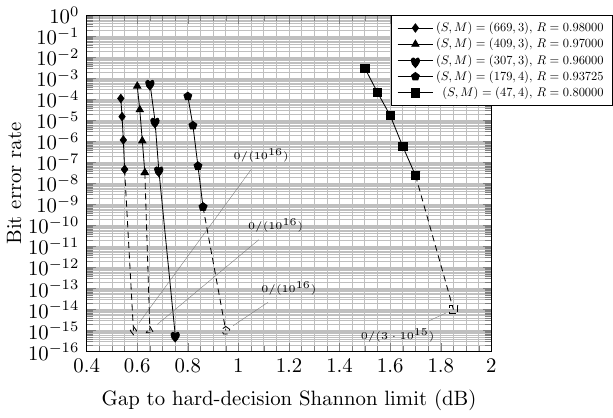}
  \caption{Simulation results for parameters in Table \ref{sim-param-table}.}\label{sim-curves}
\end{figure}

A highly-optimized software-based simulator
capable of achieving simulated throughputs of several gigabits
per second per core of a modern multi-core consumer processor 
was developed and allows for direct verification of sub-$10^{-15}$
error floors.
This is enabled in part by the simplicity of syndrome decoding 
of Hamming components.
A few representative design examples are considered 
with parameters given in Table \ref{sim-param-table} and the 
corresponding simulation results given in Fig.~\ref{sim-curves}
as bit error rate (BER) curves.
We assume transmission across a binary symmetric channel (BSC)
parameterized by the equivalent gap to the
hard-decision Shannon limit under binary phase-shift keying (BPSK) transmission 
on an additive white Gaussian noise (AWGN) channel
at the respective rates $R$ of the codes considered. 
The last two columns of Table \ref{sim-param-table}
provide the gap to the hard-decision Shannon limit
along with the corresponding 
BSC crossover probability or input BER 
for the last point of each curve, i.e., the point with least (output) BER.
Where zero bit errors were measured, a dashed line is 
used in Fig.~\ref{sim-curves} to indicate the simulated operating point
with the number of error-free bit transmissions also indicated. 
For all but the last code considered,
we simulate the transmission of at least $10^{16}$
bits at the last plotted operating point, providing
strong confidence that BERs are below $10^{-15}$.
Relative to codes using two triple-error-correcting BCH components
per bit in \cite{sukmadji,smith,Low-Rate-Hardware-Staircase},
the proposed codes perform roughly $0.2$ to $0.5$ dB worse
in terms of the gap to the hard-decision Shannon limit
as the code overhead ranges from $2\%$ to $25\%$.
In return, we expect significantly reduced power consumption.

\begin{table}[htb]
	\centering \caption{Simulation parameters.}
	\begin{tabular}{c|c|c|c|c|c|c|c|c|c|c}
		$S$ & $M$ & $r$ & $F$ & $W$ (blocks) & $W$ (Mbits) 
		& $R_\mathsf{nominal}$ & $R$ & \small{iterations} & gap (dB) & \small{input BER}\\
		\hline
		669 & 3 & 13 & 725  & 21 & 9.399 & 0.98057 & 0.98000 & 3 & 0.585 & 9.86e-4\\
		409 & 3 & 12 & 926  & 21 & 3.513 & 0.97066 & 0.97000 & 3 & 0.650 & 1.57e-3\\
		307 & 3 & 12 & 885  & 21 & 1.979 & 0.96091 & 0.96000 & 4 & 0.750 & 2.09e-3\\
		179 & 4 & 11 & 1634 & 36 & 1.153 & 0.93855 & 0.93725 & 4 & 0.950 & 3.25e-3\\
		 47 & 4 & 9  & 912  & 48 & 0.106 & 0.80851 & 0.80000 & 6 & 1.850 & 1.05e-2\\
	\end{tabular}\label{sim-param-table}
\end{table}

\FloatBarrier


\end{document}